# Influence of image segmentation on one-dimensional fluid dynamics predictions in the mouse pulmonary arteries

Mitchel J. Colebank[1], L. Mihaela Paun[2], M. Umar Qureshi[1], Naomi Chesler[3], Dirk Husmeier[2], Mette S. Olufsen[1], Laura Ellwein Fix[4*]

[1]*Mathematics, NC State University, Raleigh, NC 27695, USA*

[2]*Mathematics and Statistics, University of Glasgow, Glasgow G12 8SQ, UK*

[3]*Biomedical Engineering, University of Wisconsin-Madison, Madison, WI 53706, USA*

[4]*Mathematics and Applied Mathematics, Virginia Commonwealth University, Richmond, VA 23220, USA*

[*]**Corresponding author**

Laura Ellwein Fix

Department of Mathematics and Applied Mathematics

Virginia Commonwealth University

1015 Floyd Ave.

Richmond, VA 23220

Email: lellwein@vcu.edu

Phone: 804-828-2748

Fax: 804-828-8785

**Abstract**

Computational fluid dynamics (CFD) models are emerging as tools for assisting in diagnostic assessment of cardiovascular disease. Recent advances in image segmentation has made subject-specific modelling of the cardiovascular system a feasible task, which is particularly important in the case of pulmonary hypertension (PH), which requires a combination of invasive and non-invasive procedures for diagnosis. Uncertainty in image segmentation can easily propagate to CFD model predictions, making uncertainty quantification crucial for subject-specific models. This study quantifies the variability of one-dimensional (1D) CFD predictions by propagating the uncertainty of network geometry and connectivity to blood pressure and flow predictions. We analyse multiple segmentations of an image of an excised mouse lung using different pre-segmentation parameters. A custom algorithm extracts vessel length, vessel radii, and network connectivity for each segmented pulmonary network. We quantify uncertainty in geometric features by constructing probability densities for vessel radius and length, and then sample from these distributions and propagate uncertainties of haemodynamic predictions using a 1D CFD model. Results show that variation in network connectivity is a larger contributor to haemodynamic uncertainty than vessel radius and length.

**1 Introduction**

Definitive diagnosis of pulmonary hypertension (PH), defined as a mean pulmonary arterial blood pressure $\geq 25$ mmHg, requires a series of medical tests including invasive right heart catheterization and non-invasive computed topography (CT) imaging of the heart and lungs (1). Most diagnostic protocols interpret each data source independently to make an ultimate decision about the disease classification and severity (2), but recent studies (3,4) have proposed assimilation of haemodynamic and imaging data with computational fluid dynamics (CFD) modelling, providing insight into the structure and function of the pulmonary system. Data provided are subject to post-processing induced uncertainty, making uncertainty quantification (UQ) a vital component of the model analysis, hence the focus of this study.

Medical imaging and image segmentation have emerged as powerful non-invasive tools for disease diagnostics (5-7), providing an abundance of data for analysing the structure and function of the cardiovascular system under physiological and pathological conditions (1). Advances in image segmentation have led to semi- and fully-automated algorithms for geometric reconstruction of complex vascular regions (8,9). However, inherent uncertainty is present as most image segmentation software require manual specification of the image intensity thresholds (pre-segmentation parameters) between background and foreground. For example, van Horssen et. al (10) showed that variation in image resolution affected the cumulative volume of a cast of the coronary arterial tree after segmentation. Rempfler et. al (11) compared segmentation algorithms on retinal images, showing that posterior probability estimates for foreground pixels varied with different segmentation techniques when compared to the true segmentation or so-called "ground-truth". These two studies quantified variability in segmented networks but did not investigate how uncertainty affected CFD predictions. In contrast to the aforementioned studies, *in-vivo* images are only captured up to a finite resolution, which makes ground-truth rendering impossible.

Haemodynamic predictions (*e.g.,* cross-sectional averaged flow and pressure) in the pulmonary vasculature are often conducted using either three-dimensional (3D) (12) or one-dimensional (1D) (3) CFD models. 3D models predict local flow patterns with more precision (4) but are computationally expensive, making it difficult to perform multiple forward model evaluations for UQ (13). For instance, Sankaran et. al (14) computed 3D CFD model sensitivity to coronary stenosis diameters, using surrogate model approximations to combat high computational cost. However, they did not account for possible changes in network connectivity nor for the uncertainty from the initial segmentations of the vasculature. In contrast, 1D models are more computationally efficient, reducing the need for surrogates and allowing for investigations into variability of network connectivity. Moreover, a recent study (15) of the coronary vasculature showed that 1D models attain similar haemodynamic predictions as 3D when using appropriate boundary conditions. Recent studies have analysed 1D systemic arterial models (10,16) to understand how uncertainty in network structure impacts haemodynamics. Fossan et. al (17) devised an optimization strategy to determine the number of vessels needed to match haemodynamic predictions in the coronary arteries, and Huberts et al. (13) used polynomial chaos expansion to quantify the sensitivity of flow predictions to variations in vessel radius, informed by literature values. In contrast to the systemic circulation, the pulmonary system differs significantly, as the pulmonary vasculature is more compliant and branches more rapidly, indicating that results from the systemic circulation may not be valid for comparison.

In this study, we examine how pre-segmentation parameters impact estimated vessel radius, vessel length, and network connectivity and propagate this uncertainty to haemodynamic predictions in the pulmonary circulation. To do so, we analyse multiple segmentations of a microcomputed tomography (micro-CT) image from a mouse pulmonary arterial tree. We propagate this uncertainty using a 1D CFD model by constructing the model domain from each segmentation. We perform inverse UQ by estimating probability density functions (PDFs) for vessel radii and length, and then propagate uncertainties (forward UQ) using Monte Carlo sampling. Uncertainty in haemodynamic predictions is quantified by analysing three sets of predictions (depicted in figure 1); 1) haemodynamic predictions using 25 segmented networks (*total variation*); 2) predictions from a representative network when drawing realizations of length and radius perturbations with fixed connectivity (*parameter variation*); and 3) predictions from the same representative network when geometric parameters are fixed, but connectivity and network size are varied (*network variation*). We argue that UQ is an essential component of the model analysis when computational models are integrated into clinical protocols. The animal dataset used here (18,19) serves as a preliminary step in understanding disease progression and has potential for extrapolation to human PH.

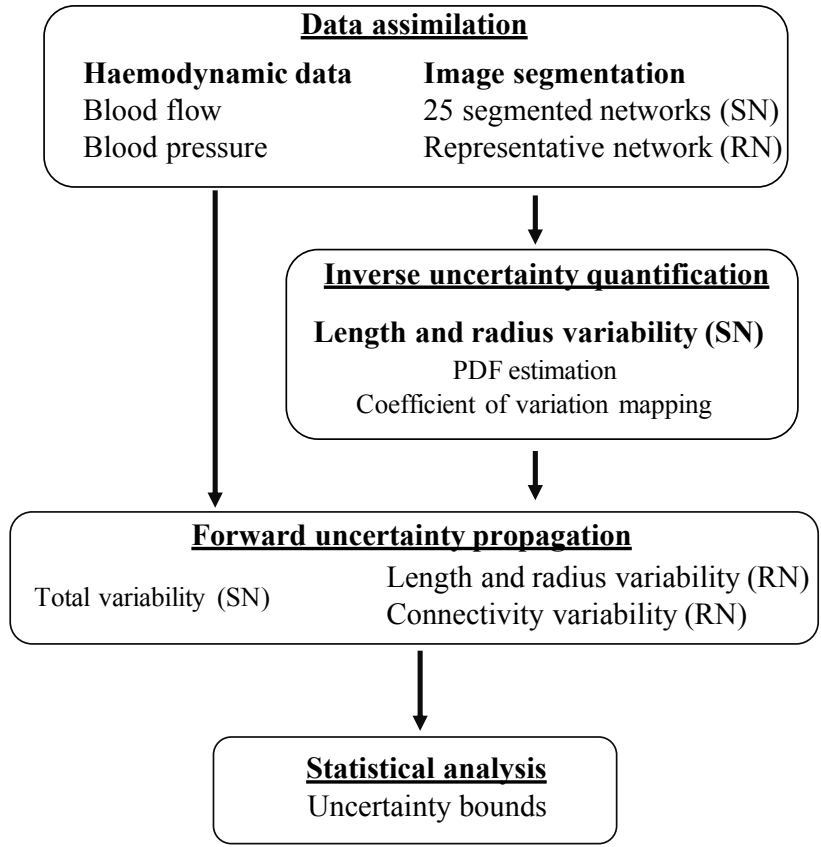

**Figure 1:** Workflow for uncertainty quantification of haemodynamics. Multiple segmentations are performed to construct the segmented networks (SNs), of which one network is selected as the represented network (RN). Inverse uncertainty quantification (UQ) is performed on the 25 SNs by estimating probability density functions (PDFs) for vessel radius and length. The 25 SNs are used in model simulations to understand the total variation, while the PDFs for the vessel dimensions are used to propagate uncertainty in the parameter variation study. Lastly, we change the structure of the RN to understand the variation induced by network connectivity. Pressure and flow predictions are then compared from the three sources of variation.

**2. Materials and methods**

**2.1 Experimental data**

This study uses existing micro-CT and haemodynamic data from two male C57BL6/J control mice aged

10-12 weeks. A detailed description of experimental protocols for the imaging and haemodynamic data can be found in Vanderpool et. al (18) and Tabima et. al (19), respectively. Briefly, haemodynamic data includes a flow waveform ensembled over 20 cardiac cycles measured using an in-line flow meter (Transonic Systems, Ithaca, NY) in the main pulmonary artery (MPA). The imaging data is obtained after euthanisation and inflation of the mouse lung at 17.2 mmHg. A cannula with outer diameter 0.127 cm and inner diameter of 0.086 cm is attached to the MPA before 360-degree imaging and reconstruction to DICOM 3.0 files. Both procedures were approved by the University of Wisconsin-Madison Institutional Animal Care and Use Committee.

## 2.2 Network reconstruction

### 2.2.1 Image segmentation

The micro-CT image is stored as a DICOM 3.0 file with voxel dimensions $497 \times 497 \times 497$. The gray-scale image (shown in figure 4a) is transformed to a binary map identifying the vascular ('foreground') and non-vascular ('background') regions using global thresholding and image segmentation in ITK-SNAP (20). Global thresholding is a pre-segmentation technique requiring a priori selection of thresholds to specify the image intensity bounds of the foreground. Threshold bounds are traditionally selected in an *ad hoc* manner to ensure that the foreground is captured (3,21,22). In addition, ITK-SNAP requires specification of a smoothing parameter to determine the boundary between the foreground and background (see figure 2). Due to the experimental protocol and use of perfused contrast, the image segmented in this study does not contain high intensity voxels from other anatomical features (*e.g.,* veins, the heart, or spine) within the region of interest. Therefore, only the lower threshold ($\theta_1$) and smoothing ($\theta_2$) pre-segmentation parameters require specification.

Acceptable intervals for ($\theta_1, \theta_2$) are determined to preserve the foreground for the large vessels across segmentations. To study segmentation induced uncertainty, we assume a uniform distribution for the two parameters, with a lower threshold range of $20 \leq \theta_1 \leq 45$ and a smoothing parameter range $3 \leq \theta_2 \leq 8$, and draw 25 realisations of pre-segmentation parameter sets ($\theta_1, \theta_2$) (given in table 1) using the *rand* function in MATLAB (Mathworks, Natick, MA). As shown in figure 3, the foreground for distal vascular segments changes significantly when ($\theta_1, \theta_2$) are varied, but maintains features for the large, proximal vessels.

We use active contour evolution, a semi-automated segmentation algorithm available in ITK-SNAP, to segment the micro-CT image. We consistently use 2000 iterations of the contour evolution, ensuring that the largest arteries carrying the majority of the blood volume are captured. The imaging protocol described in Vanderpool et. al (18) has a spatial resolution between 30-40 $\mu$m, providing a lower bound of 40$\mu$m for the measurement uncertainty diameter (20$\mu$m for radius).

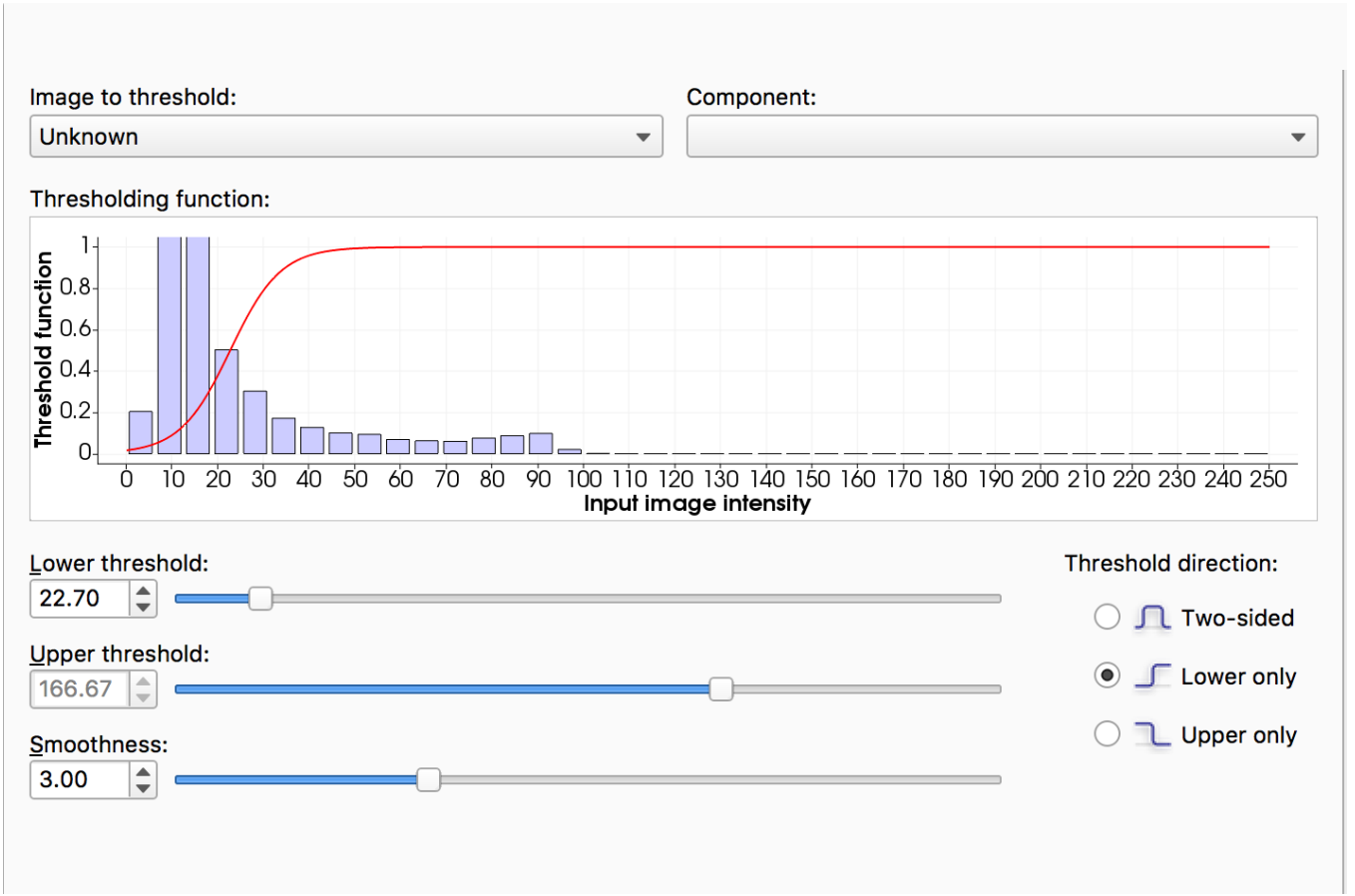

**Figure 2:** ITK-SNAP interface for prescribing $(\theta_1, \theta_2)$. Voxel intensities in the histogram are mapped to foreground and background based on thresholding function (red curve) and pre-segmentation parameters. Here, we only assume a lower threshold on image intensities, as shown by the constant value of 1 in the threshold function for all values greater than the lower threshold.

### 2.2.2 Network reconstruction

Segmented geometries are exported as surface meshes and converted to VTK polygonal files using Paraview (23) (Kitware, Clifton Park, NY). We developed a custom MATLAB algorithm to extract the network connectivity and identify all the vessels in each network. Subsequently, we use a recursive algorithm to construct a connectivity matrix identifying the geometry of the tree, which is then used in the $\theta_1 = 26, \ \theta_2 = 3.0$ $\qquad$ $\theta_1 = 36, \ \theta_2 = 3.0$ $\qquad$ $\theta_1 = 46, \ \theta_2 = 3.0$

$\theta_1 = 36, \ \theta_2 = 1.5$ $\qquad$ $\theta_1 = 36, \ \theta_2 = 3.0$ $\qquad$ $\theta_1 = 36, \ \theta_2 = 5.0$

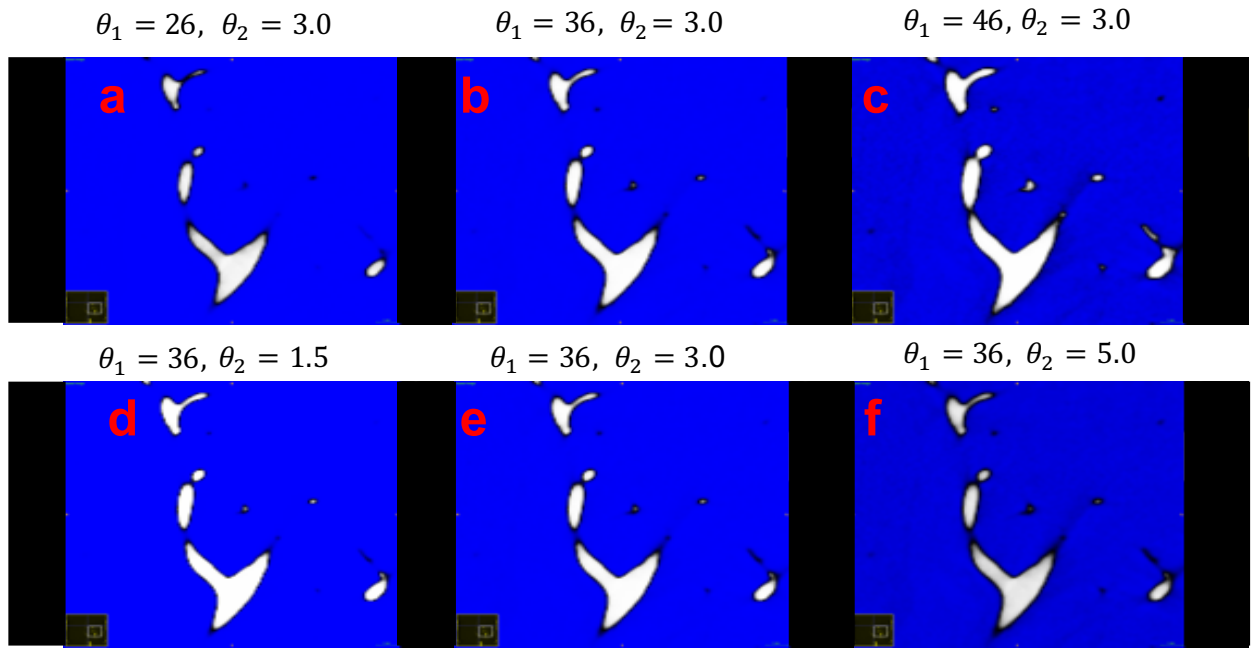

**Figure 3:** Qualitative differences in foreground (white) of distal vascular segments when changing the lower threshold ($\theta_1$) and the smoothing parameter ($\theta_2$). Top: changes in foreground with $\theta_1$; bottom: changes in foreground with $\theta_2$.

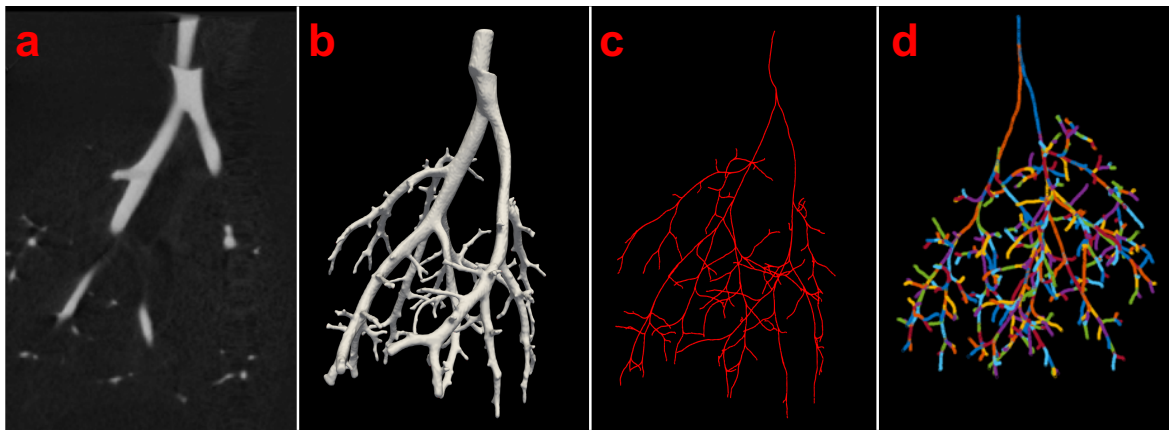

**Figure 4:** Image to network workflow. a) the foreground visible in the image file; b) the 3D rendering of the vascular foreground; c) centerlines obtained using VMTK; d) a graph representation of the network used in the 1D model with vessels (edges) and bifurcations (nodes) identified using custom MATLAB algorithms.

1D model (see supplement for algorithm detail). Figure 4 illustrates how the micro-CT image is segmented to form the 3D structure and later reduced and translated into a connected network.

A scaling factor converts voxel measurements to cm by relating voxels in the MPA to the known diameter of the cannula (0.086 cm). This scaling factor translates length and radii measurements in the entire network (18) before saving the vessel dimensions. The vessel length is calculated as the sum of the Euclidean distances between successive spatial points along each vessel. However, vessels with length less than the spatial resolution of the numerical solver ($2.5 \times 10^{-3}$ cm) are augmented to satisfy the Courant-Friedrichs-Lewy (CFL) condition of the numerical solution scheme (24). Figure 5 shows an example network and the radii estimates at each point along the network and within a single vessel. Measurements for radii vary substantially within each vessel, hence vessel radius is fixed to be the mean over the centre 80% of the individual estimates. This ensures that the ostium regions at either end of the vessel do not skew radii estimates. The MPA radius is estimated using measurements distal to the cannula before the left (LPA) and right (RPA) pulmonary arterial bifurcation.

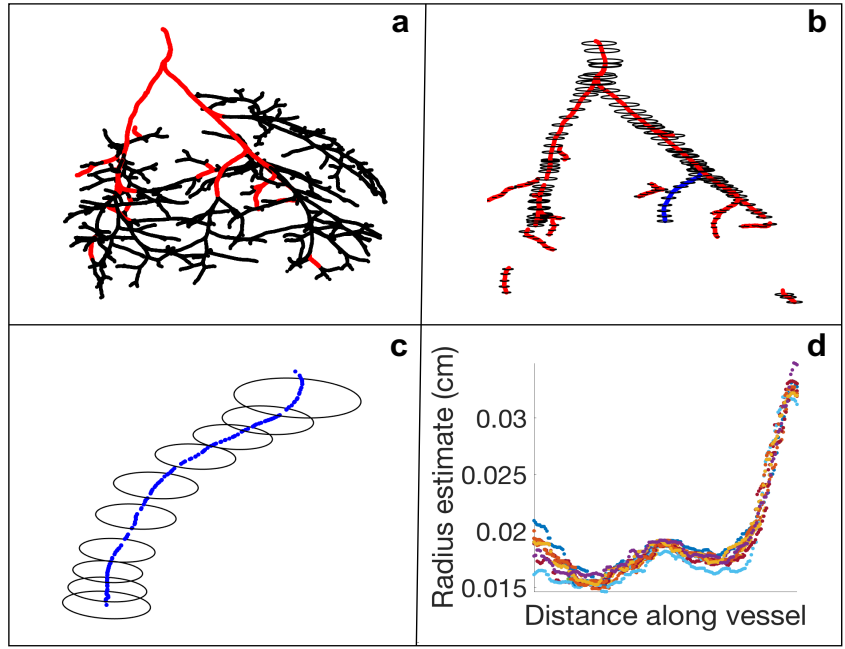

**Figure 5:** Components of an arterial tree. a) example network; b) 32 representative vessels of varying caliber identifiable in all 25 segmented networks; c) magnification of the representative vessel in blue from panel (b) depicting circles from which radius estimates are obtained; d) radius estimates along the representative vessel in panel c, where the center 80% of points are used to calculate the mean radius. Actual radius values obtained in d are calculated at orthogonal centerline slices in VMTK, while panel c shows non-orthogonal radii estimates for sake of illustrating differences in radii predictions.

We construct a connected graph using the centreline data and impose a connectivity matrix linking vessels, represented by their length and radius, and bifurcations. In addition, we capture global network features including the number of vessels, number of bifurcations (*i.e.* generations), and the total vascular volume. The CFD model used for haemodynamic assumes a binary structure, with each generation of the tree being formed by a new set of vessels.

### 2.3 Haemodynamics modelling

### 2.3.1 Blood flow model.

Similar to previous studies (3,25), we use a 1D CFD model to predict time-varying flow, pressure, and area in each vessel. The model equations for mass conservation and balance of momentum, described in detail in (3), take the form

$$\frac{\partial A}{\partial t} + \frac{\partial Q}{\partial x} = 0, \tag{2.1}$$

$$\frac{\partial Q}{\partial t} + \frac{\partial}{\partial x}\left(\frac{Q^2}{A}\right) + \frac{A}{\rho}\frac{\partial P}{\partial x} = -\frac{2\pi \nu r Q}{\delta A}, \tag{2.2}$$

respectively, where $x$ (cm) and $t$ (s) denote the axial and temporal coordinates, $A(x,t)$ (cm$^2$) denotes the cross-sectional area, $Q(x,t)$ (cm$^3$/s) the volumetric flow rate, $P(x,t)$ (mmHg) is the transmural blood pressure, and $r(x,t)$ (cm) the vessel radius. The blood density $\rho = 1.057$ (g/cm$^3$) and the kinematic viscosity $\nu = 0.0462$ (cm$^2$/s) are assumed constant (26,27). Moreover, we assume a flat velocity profile with a linearly decreasing boundary layer with thickness $\delta = \sqrt{\nu T / 2\pi}$ (cm), where $T$(s) is the length of the cardiac cycle extracted from data (3,28). To close the system of equations, we consider a linear state-equation (3,29,30) given by

$$P - P_0 = \frac{4}{3}\beta\left(1 - \sqrt{\frac{A_0}{A}}\right), \tag{2.3}$$

where $\beta = Eh/r_0 = 37.5$ mmHg describes the arterial stiffness, $E$ (mmHg) is the Young's modulus in the circumferential direction, $h$ (cm) the wall thickness, and $A_0 = \pi r_0^2$ (cm$^2$) is the reference area obtained at the reference pressure $P_0$ (mmHg). The system (2.1) - (2.3) is solved using the two-step Lax-Wendroff finite difference scheme in C++ (25).

### 2.3.2 Inflow, outflow and junction conditions

The system governed by equations (2.1) - (2.3) is hyperbolic with characteristics pointing in opposite directions, thus two boundary conditions are needed at each vessel inlet and outlet. At the network inlet (the MPA) we prescribe a measured flow waveform from a single cardiac cycle. At network bifurcations we impose two conditions ensuring conservation of flow and a continuity of pressure, giving

$$Q_p(l_p, t) = Q_{d_1}(0, t) + Q_{d_2}(0, t), \qquad P_p(l_p, t) = P_{d_1}(0, t) = P_{d_2}(0, t), \tag{2.4}$$

where the subscripts $p, d_1, d_2$ indicate the parent and daughter vessels and $l_p$ denotes the length of the parent vessel. Lastly, we impose a three element Windkessel model at the outlet of terminal vessels (4) to characterize the downstream vasculature, which relates pressure and flow via an RCR circuit model

$$\frac{dP(l,t)}{dt} = R_1 \frac{dQ(l,t)}{dt} + Q(l,t)\left(\frac{R_1 + R_2}{R_1 R_2}\right) - \frac{P(l,t)}{R_2 C_T} \tag{2.5}$$

where $R_1$ is the proximal resistance, $R_2$ is the distal resistance, and $C_T$ is the total compliance (29,31).

### 2.3.2 Parameter values

The network can be described by two sets of parameters: those attributed to the geometry (*e.g.*, vessel length, radius, and connectivity) and those attributed to the haemodynamics (viscosity, density, wall stiffness, and boundary conditions). We assume that inflow, viscosity, density, and wall stiffnes mmHg are fixed and independent of the network geometry (3,28,32) , as the objective of this study is to analyse uncertainty associated with changes in pre-segmentation parameters. Parameters remaining are those specifying the vessel radius, vessel length, and Windkessel outflow boundary conditions ($R_1, R_2, C_T$), which depend on the network structure (3,17).

For each network, vessel radii and length are determined from the segmentation, while estimates are needed for Windkessel parameters. Similar to our previous study (3), we assume that the total compliance $C_T$ can be determined from the time constant $\tau = R_T C_T$, where $R_T = R_1 + R_2$ is the total vascular resistance (3). $R_T$ is computed as the ratio of mean pressure to mean flow, *i.e.* $R_T = \overline{P}/\overline{Q}$, and as discussed in our previous studies (3,30), *a priori* resistance values for each terminal vessel can be calculated using  Poiseuille's equation, relating mean pressure and flow via the vessel dimensions. Both junction conditions in equation (2.4) are used together with Poiseuille's law to give the mean flow distribution relationship

$$\overline{Q}_{d_1} = \overline{Q}_p \frac{\xi_{d_1}}{\xi_{d_1} + \xi_{d_2}}, \quad \text{and} \quad \overline{Q}_{d_2} = \overline{Q}_p \frac{\xi_{d_2}}{\xi_{d_1} + \xi_{d_1}}, \tag{2.6}$$

where $\xi_i = r_i^4 / l_i$, consistent with Poiseuille's equation (see the supplement for details). Finally, we set $R_1 = 0.2 R_T$ and $R_2 = 0.8 R_T$ (3).

### 2.4 Inverse uncertainty quantification

We employ inverse UQ to estimate vessel length and radius PDFs over the 25 segmented networks. To compare measurements across segmentations, PDFs are computed for radius and length from a 32-vessel subset after data standardization. Two estimation techniques, kernel density estimation (KDE) and Gaussian process (GP) density estimation, are used to compare estimated PDFs. Weighted least squares regression and GP regression are used to remedy the issues of non-constant variance, *i.e.* heteroscedasticity, in vessel dimensions.

### 2.4.1 Data standardization

A subset of 32 pulmonary vessels of various calibre (see figure 5a) is selected from the 25 segmented networks. The 32 vessels are visible in all 25 networks and contain radius and length measurements that encompass the full range of measurements in the networks. Length and radius measurements are standardized using

$$s_{i,j}^* = \frac{s_{i,j} - \bar{s}_i}{\sigma_{s_i}}, \tag{2.7}$$

where $s_{i,j}$, $s = r, l$ are the measured quantities from the $i$th vessel and $j$th segmentation, and $\bar{s}_i$ and $\sigma_{s_i}$ are the mean and the standard deviations of these quantities across the 25 networks.

### 2.4.2 Density estimation

We used KDE, a nonparametric technique (33), to estimate the PDFs for radius and length. These techniques require specification of a bandwidth parameter, determining how influential each data point is in the density estimation. We consider both Silverman's rule of thumb (33) and maximum likelihood leave-one-out cross validation (34) for bandwidth estimation. These methods are compared to logistic GP density estimation (35) using the *GP Stuff* toolkit in MATLAB (36). Due to space restrictions, the methodological details have been relegated to the supplementary material, §S.5.

### 2.4.3 Statistical models for computing the length and radius variance

The PDFs constructed from the 32-vessel subset are representative of the overall variation in the length and radius across all the segmented networks. However, the magnitude of $\sigma_{l_i}$ and $\sigma_{r_i}$ vary from vessel to vessel and need to be modeled explicitly before performing forward UQ. The coefficient of variation, $c_v^{s_i} = \sigma_{s_i} / \bar{s}_i$, is used for comparing the length and radius measurements to their variability.

The statistical model $\phi(\bar{s}_i) = c_v^{s_i}$ relates the average measurements of radius and length across segmentations to their coefficient of variation. The variance of the measurements exhibits heteroscedasticity, as smaller vessel segments are more sensitive to pre-segmentation parameters leading to non-constant variance. This violates the assumptions of ordinary linear regression; hence we consider weighted least squares regression and GP regression with input-dependent noise (37).

Traditional deterministic weighted least squares regression iteratively fits regression models by updating weights for each data point. The optimal weights (optimal in a maximum likelihood sense) are given by the inverse of the variance of the response $\phi(\bar{s}_i)$ (38). Since this variance is unknown, we approximate it by $w_i = 1/\epsilon_i^2$, where $\epsilon_i$ is the residual from the unweighted regression model, reducing the impact of highly variable observations on the regression prediction. We consider exponential, logarithmic, square root, and linear weighted least squares regression models. For GP regression, we employ two GPs for the response, $c_v^{s_i}$, and the latent variance of $c_v^{s_i}$. The GPs used the Matérn covariance function (39) with a smoothness parameter $\nu = 5/2$ (see the supplement for more details).

### 2.5 Forward Uncertainty Quantification

Forward UQ propagates model and parameter uncertainties to simulated quantities of interest. An issue here is that both the network (the number of vessels and connectivity) and model parameters (length, radius, and boundary conditions) give rise to uncertainty. To analyse the posterior variation in model predictions, we pursue three sets of simulations determining (i) the total variation of haemodynamic predictions associated with segmentation, (ii) the variation to changes in model parameters, and (iii) the variation to network size and connectivity. The first set of simulations (i) use the 25 segmented networks, whereas the last two (ii-iii) are conducted in the representative network.

### 2.5.1 Total variation

We evaluate the CFD model using each of the 25 segmented networks to quantify the total variation of flow and pressure predictions in the MPA, LPA, and RPA. The variation observed is attributed to several sources of uncertainty, including the parameters of the model and the size and connectivity of the network. Once the total variation is calculated, we quantify the relative contributions from the parameter and network variation.

### 2.5.2 Representative network

We determine the representative network by first computing the pressure waveform in the MPA for each of the 25 segmented networks and then calculate the least squares cost between the waveform and the ensemble averaged waveform from all 25 networks. The network with the smallest least squares cost is designated the representative network and used to determine (ii) the parameter variation and (iii) the network variation.

### 2.5.3 Parameter variation

As mentioned in §2.3, we assume that density, viscosity, and vessel stiffness are constant while parameters impacted by image segmentation, including vessel length, radius, and boundary conditions, vary. The outflow boundary conditions are dependent on vessel length and radius; thereby, we analyse the variation in model predictions associated with changes in vessel dimensions. We conduct the computations in the representative network and explicitly study what part of the variation is attributed to these model parameters.

We compute inverse cumulative distribution functions (CDFs) for the length and radius PDFs. The inverse CDF $F_s^{-1}(\alpha)$ is a nondecreasing function defined on the interval $[0, 1]$ that provides values from the original PDF, allowing for inverse transform sampling for forward UQ (13). Briefly, let $u$ be a realization from a uniform distribution $u \sim \mathcal{U}(0,1)$, and define the realization from the inverse CDF as $F_s^{-1}(u)$. There exists a mapping from the realization to the inverse CDF for the radius and the length via $\gamma_r = F_r^{-1}(u)$ and $\gamma_l = F_l^{-1}(u)$. We draw samples from the inverse CDF to provide standardized measurements $l^*$ and $r^*$ for length and radius.

We then define a mapping from the inverse CDF of $\bar{s}_i$ in vessel $i$ to the perturbed values $\hat{s}_i$ (in units of cm). We write $F_s^{-1}(u) = (\hat{s}_i - \bar{s}_i)/\sigma_{s_i}$ and rewrite the standard deviation as $\sigma_{s_i} = c_v^{s_i} \cdot \bar{s}_i = \phi(\bar{s}_i) \cdot \bar{s}_i$, where $\phi(\bar{s}_i)$ is the statistical model found from §2.4.3. This gives

$$\hat{s}_i = (F_s^{-1}(u) \cdot \phi(\bar{s}_i) + 1) \cdot \bar{s}_i \tag{2.8}$$

for each average measurement $\bar{s}_i$ in vessel $i$. The values $\hat{s}$ are used as the dimensions for each vessel in the 1D model when doing the forward UQ. We propagate uncertainties in the representative network by setting the average measurement $\bar{s}_i = s_i^{rep}$, where $s_i^{rep}$ are the original measurements from the representative network. To ensure convergence in the posterior of the haemodynamics (29), we draw $M = 10^4$ realizations using Monte Carlo sampling to perturb the length and radius values. The pseudo algorithm for UQ propagation is given as follows:

1. Draw a random sample $u \sim \mathcal{U}(0,1)$.
2. Map the sample to $F_r^{-1}(u)$ and $F_l^{-1}(u)$.
3. Perturb the nominal radius and length by using equation (2.8).
4. Run 1D CFD model with new radius and length values.
5. Repeat steps 1-4 $M$ times.

### 2.5.4 Network variation

We alter the total network size and connectivity by fixing vessel dimensions and instead varying the number of vessels representative network used in §2.5.2. The smallest terminal vessels in the tree contain the least number of voxels and are thus the most susceptible to changes in pre-segmentation parameters. We simulate the possible exclusion of vessels by systematically eliminating terminal vessel pairs by first calculating the total volume of each terminal vessel (*i.e.* $V_{tot} = \pi r^2 l$) and then removing the terminal pair with the smallest volume. The truncation begins at the smallest terminal vessels and continues until only the MPA, LPA, and RPA remain. While reducing the network size, we ensure that the total resistance, total compliance and the total mean flow in the network is conserved. This is an important step in isolating the effects of Windkessel parameters from the geometric parameters, as these quantities dictate the calculated resistance and compliance estimates each terminal vessel.

### 3 Results

We analyse the total variation of flow and pressure predictions and identify the relative contributions from variations in model parameters and in the network. The total variation in the model predictions, attributed to changes in vessel length and radius as well as network size and connectivity, is quantified by comparing simulations in the MPA, LPA, and RPA using each of the 25 networks. We compare this with simulations corresponding to the variation in vessel radius and length and variation in network structure in a representative network.

### 3.1 Network statistics

Figure 6 summarizes network characteristics obtained from the 25 segmented networks. Total cross-sectional area, defined as the average and sum of cross-sectional areas in each generation, as well as the average cross-sectional area and number of vessels in each generation, are shown in figure 6. The average number of vessels in the network is 437 with a standard deviation of 76 and the mean number of generations across segmentations is approximately 17. The number of vessels and total cross-sectional area of the networks are relatively consistent across segmentations up until the 6th generation, after which the results deviate. Most  segmentations achieve a maximum number of vessels and cross-sectional area between generations 8 and 14, while the average cross-sectional area rapidly decreases until the 5th generation, and then remain fairly constant afterward. Analysis across all networks in figure 6d shows that one network (corresponding to $(\theta_1, \theta_2) = (44, 7.6)$) is an outlier, having significantly fewer vessels and a lower total cross-sectional area. Table 1 includes all pre-segmentation parameter sets used in the repeated segmentations as well as network level features.

### 3.2 Inverse UQ

Figure 7 shows the estimated length and radius densities for the 32 representative vessels using KDE with bandwidths calculated via Silverman's rule and maximum likelihood cross validation, as well as densities obtained using GPs (see figure 5). The standard deviation for each of the 32 vessels are used to standardize the data points (see equation (2.7)) before applying density estimation techniques. The maximum coefficient of variation across all 32 vessels is 21% for the radius and 49% for the length estimate. The bandwidth estimates for Silverman's rule are $H_l^S = 2.038 \times 10^{-1}$ and $H_r^S = 1.573 \times 10^{-1}$ while the estimated bandwidths using the maximum likelihood cross validation are $H_l^{MLCV} = 1.808$ and $H_r^{MLCV} = 6.887 \times 10^{-1}$ for the length and radius densities, respectively. Silverman's rule bandwidth shows overfitting, while the KDE using the maximum likelihood cross validation bandwidths over-smooth the density relative to the GP. These results suggest that the GP is the $H_r^{MLCV} = 6.887 \times 10^{-1}$ for the length and radius densities, respectively. In general, the KDE with the best density approximation, and it is therefore chosen for the forward uncertainty propagation in §3.3.

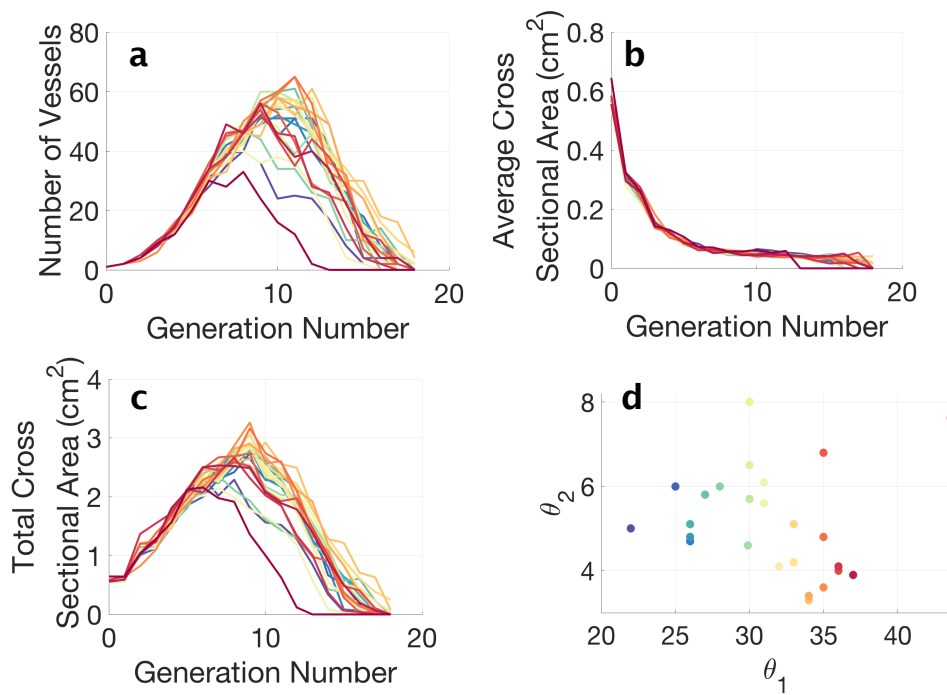

**Figure 6:** Morphometric features from the 25 segmentations marked by different colored lines. The number of vessels (a) is consistent between segmentations until the 5$^\text{th}$ generation. The average cross-sectional area (b) decreases rapidly after the 1$^\text{st}$ generation, while the total cross-sectional area (c) varies significantly between segmentations. The segmentation parameters are plotted against each other in (d), with a clear outlier present at $(44, 7.6)$ (in pink) indicating a set of pre-segmentation parameters that have marked effects on the network structure. The outlier has lower number of vessels and total cross-sectional area as depicted in the pink curve in panels (a) and (c).

Weighted least squares with exponential, logarithmic, square root, and linear regression functions are unable to resolve the heteroscedastic nature of the data (plots not shown). We use the GP regression model with input dependent noise to construct an estimate of $\phi(\bar{s}_i)$, resolving the issue of heteroscedasticity. Figure 8 panels a and b show the GP regression for $c_v^{r_i}$ and $c_v^{l_i}$, respectively, while panels c and d depict the latent variance. The coefficient of variation for vessel measurements across segmentations increases as the measurements decrease. The variance for $c_v^l$ increases as the length decreases, yet the variance of $c_v^r$ has a sharp decrease in radius measurements. Both GP models stay above the minimum variability of $20\mu m$.

**3.3**    **Table 1.** *Summary of pre-segmentation parameters and network features*

| Pre-segmentation parameters $(\theta_1, \theta_2)$ | Number of vessels | Number of generations | Number of terminal vessels | Total volume $(cm^3)$ |
|---|---|---|---|---|
| (22, 5.0) | 276 | 15 | 149 | 21.0871 |
| (25, 6.0) | 422 | 17 | 226 | 21.3407 |
| (26, 4.7) | 415 | 17 | 219 | 22.3524 |
| (26, 4.8) | 425 | 18 | 227 | 22.8591 |
| (26, 5.1) | 441 | 17 | 234 | 22.7031 |
| (27, 5.8) | 450 | 17 | 240 | 22.9599 |
| (28, 6.0) | 333 | 15 | 178 | 20.6542 |
| (30, 4.6) | 428 | 16 | 230 | 21.7283 |
| (30, 5.7) | 461 | 17 | 245 | 23.0039 |
| (30, 6.5) | 476 | 18 | 252 | 23.1922 |
| (30, 8.0) | 409 | 16 | 220 | 21.7642 |
| (31, 5.6) | 462 | 18 | 246 | 23.3346 |
| (31, 6.1) | 310 | 15 | 164 | 18.2311 |
| (32, 4.1) | 419 | 16 | 220 | 22.2851 |
| (33, 4.2) | 446 | 18 | 239 | 23.0664 |
| (33, 5.1) | 505 | 18 | 269 | 24.6089 |
| (34, 3.3) | 495 | 18 | 265 | 24.1804 |
| (34, 3.4) | 474 | 17 | 257 | 24.2923 |
| (35, 3.6) | 459 | 17 | 242 | 23.2488 |
| (35, 4.8) | 470 | 17 | 250 | 23.0868 |
| (35, 6.8) | 404 | 17 | 214 | 22.7536 |
| (36, 4.0) | 419 | 17 | 226 | 22.0391 |
| (36, 4.1) | 376 | 16 | 197 | 22.5833 |
| (37, 3.9) | 409 | 17 | 221 | 21.6596 |
| (44, 7.6) | 185 | 12 | 98 | 20.4368 |

## Forward UQ

The MPA flow data is used as an inflow boundary condition and as a result does not change in any of the simulations. The ensemble averaged pressure predictions in the MPA, LPA, and RPA along with $\pm$ two standard deviations are shown in the first column of figure 9. Mean, diastolic and pulse pressure and max flow, min flow, and total volume, are given in table 2. The flow distribution to the LPA is much larger than the RPA, a consequence of the larger radius of the LPA that allows for greater fluid flow. The ensemble averaged pressure waveform calculated from the 25 networks identifies the network generated by $(\theta_1, \theta_2) = (33, 5.1)$ as the representative network.

For the parameter variation component of the study, we use the inverse sampling methodology defined in §2.5.3 to propagate $10^4$ realizations of perturbed radius and length values in the representative network. The second column of figure 9 shows the model predictions along with the mean and $\pm$ two standard deviations from the mean. The variation in the MPA, LPA, and RPA systolic and pulse pressure predictions are significantly larger than the those observed in the mean and diastolic pressures (see table 2). The flow predictions in the LPA and RPA have larger variability with respect to the mean and max flow in comparison to the minimum flow.

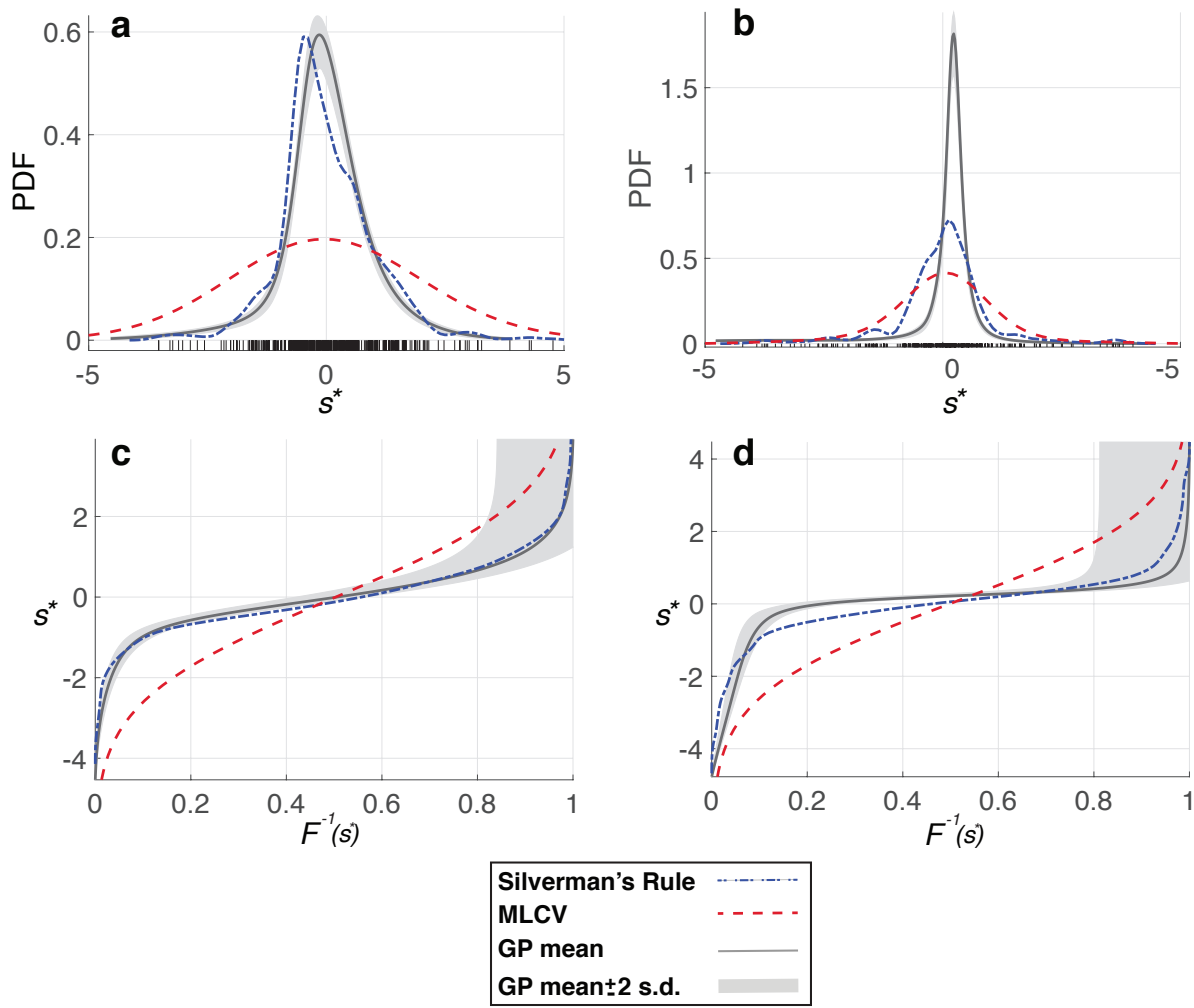

**Figure 7**: Density Estimates (a) and (b) and inverse cumulative distribution functions (c) and (d) for the standardized radius and length values, respectively, measured in the 32-vessel subset. The bandwidth parameters used for the length and radius KDEs were determined using Silverman's rule (blue, dash dot) and maximum likelihood cross-validation (MLCV, red, dashed). The Gaussian process (GP) mean and 95% confidence interval are included as well (solid curve and grey bands, respectively). Standardized values are denoted by the black tick marks in panels (a) and (b).

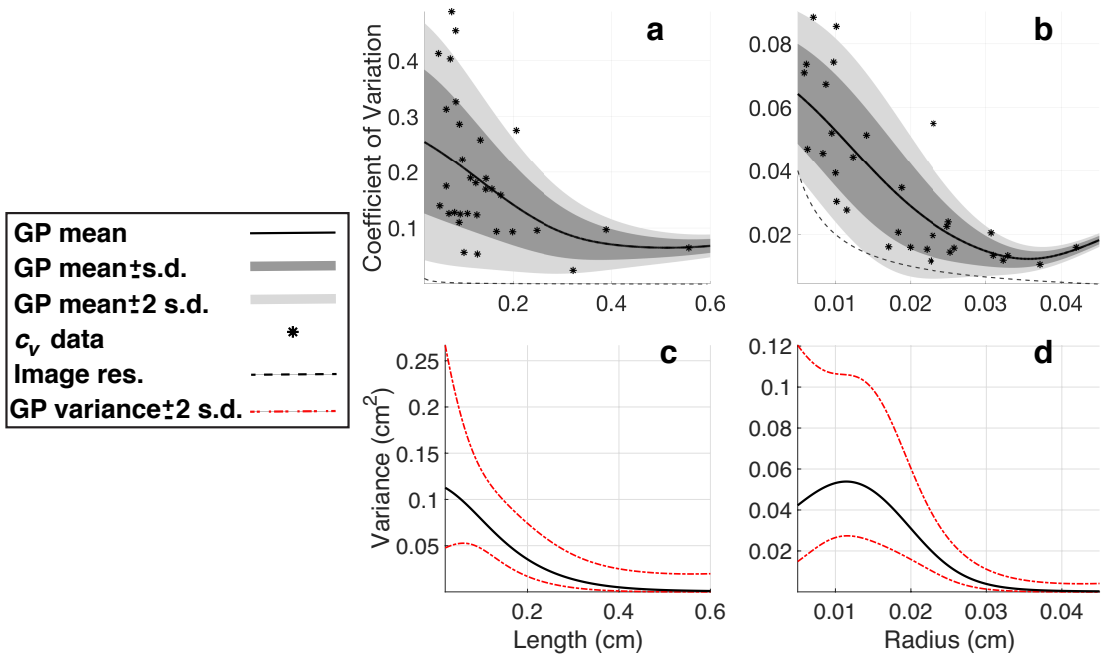

**Figure 8:** Gaussian Process (GP) regression using non-constant variance for the relationship between length and radius and their coefficient of variation ($c_v$). The GP means and standard deviations are computed from the $c_v$ data obtained from the 32-vessel subset (asterisks) and plotted against the analytical bound of the image resolution (blue, dash-dot curve). The mean of the GPs and $\pm$ one and two standard deviations (SD) from the mean are shown in (a) and (b) in black, dark grey, and light grey, respectively. The variance of the GPs in (c) and (d) are predicted using an additional GP and provide a mean (black) and variance (dashed curve) for the variance estimate. Both mean curves in (a) and (b) are above the uncertainty bound of the imaging protocol.

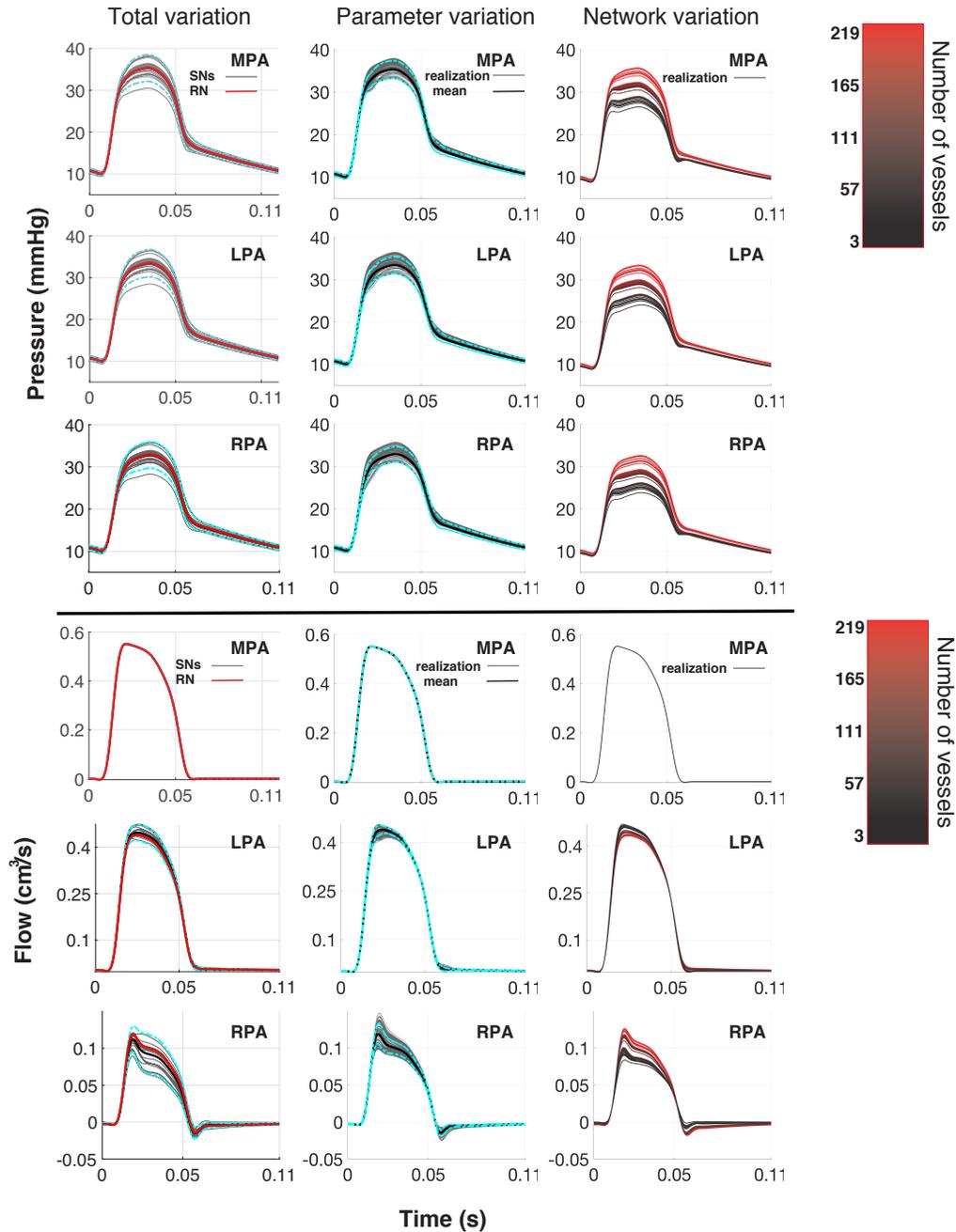

**Figure 9.** Pressure and flow predictions in the first pulmonary bifurcation when studying total variation, parameter variation, and network variation. Predictions from the total variation include simulations in the 25 segmented networks, the representative network $\pm 2$ s.d. from the mean (blue, dash-dot). The parameter variation plots (2nd column) show the 10,000 Monte Carlo realizations (grey) along with the mean (black) $\pm 2$ s.d. from the mean (blue, dash-dot). Lastly, the network variation predictions (3rd column) show the predictions when using 219 vessels in the network (bright red) up until the network is reduced to the MPA, LPA, and RPA (darkest black).

The variation attributed to network size and connectivity is calculated by fixing each vessel's radius and length in the representative network before reducing the full network iteratively. As described in §2.5.4, we reduce the network by starting at the smallest branches and moving towards the proximal vasculature while ensuring that Windkessel boundary conditions are adjusted for each simulation (see figure 9). Overall, reducing the number of vessels from 219 in the largest network to 3 in smallest network introduces a discrepancy of approximately 10 mmHg in the pressure predictions.

## 4 Discussion

We investigate three types of segmentation induced variation: the total variation arising from changes in pre-segmentation parameters, variation due to changes in vessel length and radius, and variation with respect to network connectivity and size. Results show that haemodynamic predictions vary more when changing network structure in comparison to changing haemodynamic model parameters.

### 4.1 Segmentation and construction of network graphs

Results show that pre-segmentation parameters drastically influence the number of vessels in the network, while the number of generations attainable remains relatively consistent. It is apparent that the extent of the network obtained from image segmentation is strongly linked to the range of image intensities considered in the foreground via choice of $(\theta_1, \theta_2)$.. The largest vascular tree used in this study contains 500 vessels, a small fraction of the thousands of blood vessels that comprise the full pulmonary arterial system (4,9). We expect the trends seen in figure 5 to continue if more vessels are detected by the imaging. The techniques employed here study uncertainty induced by global thresholding, but could be applied when other pre-segmentation techniques are used. Global thresholding is a commonly used technique (3,21,22), but is only one of many methods that can be used for image segmentation.

The variability in the total number of vessels for a given segmented network highlights the variation attributed to segmentation. This would be expected in other networks that exhibit dispersive branching patterns, such as the coronary arteries (10) or cerebral vasculature (31). We employed a generation based ordering scheme to describe the branching structure, where each bifurcation is considered a new generation of blood vessels. In contrast, other authors (40) have used other ordering systems, *e.g.*, Strahler (41) schemes, to identify structural properties of the pulmonary system.

### 4.2 Image-to-CFD simulation integration

**Table 2.** *Results from simulations*

| **Pressure** | | | | |
|---|---|---|---|---|
| **Total variation** | Mean pressure | Systolic pressure | Diastolic pressure | Pulse pressure |
| MPA | $20.36 \pm 0.78$ | $35.35 \pm 1.63$ | $10.02 \pm 0.27$ | $25.33 \pm 1.39$ |
| LPA | $19.66 \pm 0.79$ | $33.46 \pm 1.67$ | $10.00 \pm 0.27$ | $23.45 \pm 1.43$ |
| RPA | $19.52 \pm 0.78$ | $32.83 \pm 1.60$ | $10.10 \pm 0.28$ | $22.74 \pm 1.34$ |
| **Parameter variation** | Mean pressure | Systolic pressure | Diastolic Pressure | Pulse Pressure |
| MPA | $20.38 \pm 0.54$ | $35.37 \pm 1.03$ | $10.04 \pm 0.23$ | $25.33 \pm 0.82$ |
| LPA | $19.68 \pm 0.53$ | $33.46 \pm 0.99$ | $10.02 \pm 0.23$ | $23.43 \pm 0.78$ |
| RPA | $19.56 \pm 0.50$ | $33.46 \pm 0.90$ | $10.11 \pm 0.24$ | $22.80 \pm 0.69$ |
| **Network variation** | Mean pressure | Systolic pressure | Diastolic Pressure | Pulse Pressure |
| MPA | $18.29 \pm 0.84$ | $31.70 \pm 2.07$ | $9.08 \pm 0.18$ | $22.63 \pm 1.91$ |
| LPA | $17.44 \pm 0.86$ | $29.34 \pm 2.13$ | $9.08 \pm 0.17$ | $20.27 \pm 1.97$ |
| RPA | $17.31 \pm 0.83$ | $28.71 \pm 1.96$ | $9.15 \pm 0.20$ | $19.56 \pm 1.77$ |

| **Flow** | | | | |
|---|---|---|---|---|
| **Total variation** | Mean flow | Max Flow | Min flow | Volume |
| LPA | $0.142 \pm 0.004$ | $0.447 \pm 0.013$ | $-0.000 \pm 0.000$ | $0.016 \pm 0.000$ |
| RPA | $0.027 \pm 0.004$ | $0.113 \pm 0.009$ | $-0.015 \pm 0.004$ | $0.003 \pm 0.000$ |
| **Parameter variation** | Mean flow | Max flow | Min flow | Volume |
| LPA | $0.140 \pm 0.001$ | $0.439 \pm 0.006$ | $0.000 \pm 0.015$ | $0.015 \pm 0.000$ |
| RPA | $0.029 \pm 0.001$ | $0.119 \pm 0.007$ | $-0.014 \pm 0.002$ | $0.003 \pm 0.000$ |
| **Network variation** | Mean flow | Max flow | Min flow | Volume |
| LPA | $0.141 \pm 0.001$ | $0.447 \pm 0.009$ | $-0.001 \pm 0.001$ | $0.016 \pm 0.000$ |
| RPA | $0.027 \pm 0.001$ | $0.009 \pm 0.010$ | $-0.014 \pm 0.004$ | $0.003 \pm 0.000$ |

Values are expressed as means $\pm$ s.d.; pressure values are in units of mmHg, flow values are in units of cm$^3$/s, and volume values are in units of cm$^3$.

While previous studies (4,31) have used large networks in 1D models, this work is the first to explicitly study vessel dimension and network uncertainty in the pulmonary system. Moreover, the methodology developed herein can be used to generate a 1D model network for any vascular system. A limitation of the 1D model is that it does not consider the branching angles of the vessels, which should be investigated further, as this may also be a source of uncertainty from the segmentation. In addition, the experimental protocol inhibited the same mouse from being used for both the haemodynamic and imaging data. While this is a limitation for possible parameter inference, our methodology still captures variability in model predictions due to uncertainty in the vessel dimensions and network structure.

### 4.3 Inverse uncertainty quantification

KDEs and GPs are commonly used technique (35,42), receiving little to no attention in cardiovascular modelling. This study is the first to investigate the use of GPs in density estimation for vascular measurements. Forward UQ is typically carried out by assuming a parametric parameter distribution *a priori*, forcing prior assumptions on the unknown parameter distributions. By estimating the density directly from repeated measurements, we construct a nonparametric, representative density describing the uncertainty of the measurements across segmentations without prior assumptions.

The standardized measurements allow us to generalise the uncertainty of the 32-vessel subset to the entire vascular network, increasing the robustness of the density estimates and leading to a better representation of the distribution. As shown in figure 8, the three density estimates are similar in the mode of the distribution (approximately zero); however the GP density estimation allows for additional UQ in both the density and CDF estimates (35). We construct marginal density estimates for the PDFs of radius and length, a limitation, as this assumes independence among the two quantities. PDF estimation methods that account for correlation between radius and length measurements should be investigated further.

GP regression is necessary for the data provided as weighted least squares did not correct the heteroscedastic variance. The coefficient of variation for the measurements increased as the measured dimensions decreased in value, suggesting that smaller vessels have larger fluctuations estimated dimensions when changing pre-segmentation parameters. The gradual increase in coefficient of variation indicates that the variance of the vessel dimensions increases faster their average value. Similar conclusions have been made in simulations predicting the fractional flow reserve in coronary crowns (10), as the smaller regions of the vasculature were susceptible to higher segmentation error. However, our work is the first to consider estimated, nonparametric densities for UQ propagation, and does not require *a priori* distribution assumptions.

### 4.4 Total variation of model simulations

The total network size obtained from the segmentation procedure has several effects on the model output. As shown in table 2, changes in network topology due to segmentation induced a variation in systolic pressure that was nearly 6 times larger than the variation of diastolic pressure. Moreover, we observe that the total variation for the systolic and pulse pressure is larger in comparison to the mean and diastolic pressure. All four of these pressure metrics are typically used in diagnostic tools of diseases such as PH (2). Though systolic pressure and pulse pressure have a small standard deviation (approximately 5% relative to the mean),  studies investigating coronary related mortality found that these pressure quantities were important for risk assessment in patients with congenital heart disease (43). This further indicates a need for UQ when using these models for cardiovascular diseases diagnostics and risk assessment.

**4.5 Parameter variation**

The standard deviation of diastolic pressure was greater for radius and length perturbations than changing network size and connectivity. This suggests that changes in vessel dimensions and nominal boundary conditions can ultimately raise the diastolic pressure of the system, which is expected in the case of chronic vascular remodelling (2). Parameter variation only accounted for approximately 30% of the total variation in the pulse pressure and had less of an effect on all other pressure and flow quantities when compared to the network variation results. However, larger networks encompassing the entire pulmonary tree will increase the parameter uncertainty, as they correspond to more vessels and more uncertain estimates of radius and length. This would in turn bias haemodynamic parameter estimates, since network predictions would be based on the vessel dimensions obtained from an initial segmentation (14). While we consider uncertain measurements of radius and length, we did not account for the effects of other uncertain inputs such as the inflow profile, viscosity, or arterial stiffness, as they have been investigated elsewhere (13,30).

**4.6 Network variation**

The largest effect on pressure and flow predictions in the network are attributed to changes in network connectivity and size, as seen in figure 9. We use a Poiseuille based scheme to distribute resistance throughout the network, as described in §2.3.2, which introduces an impedance mismatch at each terminal vessel. While it is not discussed at length here, reflected pressure waves due to this mismatch become prevalent as successive vessels are added to the system, leading to an increased pressure (16,46). Other authors have considered non-reflective boundary conditions (15,44), yet it is known that wave reflections may occur in the pulmonary system when PH is present (47), illustrating the need for reflective boundary conditions in the model.

These results suggest that the size of the network used in CFD modelling can play a large role in predictions of pressure and flow. Our results show that there are three instances where reducing part of the network causes a larger change in pressure, which agree with a previous investigation by Epstein *et al.* (44) that showed a critical threshold in the number of vessels that lead to larger discrepancy in haemodynamic predictions. Moreover, changes in network size will lead to changes in optimal parameter values during parameter inference. It is often the case that haemodynamic data is only available in select locations of the vascular system (3,29,30), making the problem ill-posed as parameters describing stiffness, compliance, and vascular resistance will obtain different optimal values depending on the size of the network used in CFD simulations. This further indicates that uncertainty in the network structure must be taken into account when using 1D CFD models for clinical decision making (30).

**4.7 Future directions**

We consider 3 element Windkessel models, often used in cardiovascular parameter estimation (29,30,44), as the boundary conditions for the 1D model, yet these models lack complex physiological resistance beyond the segmented vessels. In contrast, structured tree boundary conditions (24,25,28) can provide a more physiological means for approximating downstream resistance, and attempt to capture network structure that is beyond the limits of image segmentation. In addition, future subject-specific models of the pulmonary circulation should allow for trifurcations and angles in the vascular tree, thus accounting for more of the physiological traits of the network. Future human-based studies will incorporate non-invasive flow and imaging data from the same patient in the model.

**5 Conclusions**

We have presented an in-depth study of the uncertainty that arises from subject-specific medical image geometries in 1D CFD models. Uncertainty of model predictions must be accounted for in the absence of a "true solution." This work identifies the uncertainties pertaining to image segmentation by explicitly measuring the variation in radius and length measurements of a subset of vascular segments. The propagation of geometric uncertainties through CFD models has been done previously (14,45), but this is the first time these techniques have been used in the 1D CFD framework of the pulmonary circulation. Another novelty of this work is in estimating densities of radius and length from data obtained using state-of-the-art nonparametric techniques, rather than assuming a fixed and potentially biased functional form of the distribution *a priori*. Moreover, our study is the first to perform UQ on the dimensions and network topology of a 1D CFD model in an expansive pulmonary vascular network. Our results show that the network variation has the most influence on nominal predictions of pressure and flow while changes in vessel length and radius have less impact on haemodynamic predictions.

**Funding**

This work was supported by the NSF-DMS (1246991 and 1615820), the EPSRC (EP/N014642/1), the NIH (R01 HL-086939), and the American Heart Foundation (19PRE34380459).

**Acknowledgements**

We thank Dr. Diana Tabima for her review and input on the experimental protocol.